\documentclass[pra,twocolumn,preprintnumbers,amsmath,amssymb,superscriptaddress]{revtex4-1}

\usepackage{graphicx}
\usepackage{color}
\usepackage{xcolor}
\usepackage{dcolumn}
\usepackage{bm,bbm,bbold,braket}
\usepackage{amssymb,amsfonts,amsmath}
\usepackage{amstext}
\usepackage{MnSymbol}

\usepackage[T1]{fontenc}
\usepackage[latin1]{inputenc}
\usepackage[english]{babel}

\usepackage[colorlinks=true,     linkcolor=blue,     citecolor=blue, urlcolor=blue]{hyperref}
\usepackage{relsize}
\usepackage{subfigure}
\usepackage{hhline}
\usepackage{booktabs}

\newcommand{\Eref}[1]{Eq.~\eqref{#1}}
\newcommand{\Fref}[1]{Fig.~\ref{#1}}

\newcommand{\be}{\begin{equation}}
\newcommand{\ee}{\end{equation}}
\newcommand{\ba}{\begin{eqnarray}}
\newcommand{\ea}{\end{eqnarray}}
\newcommand{\eq}[1]{\begin{equation} #1 \end{equation}}

\newcommand{\ud}{\mathrm{d}}
\newcommand{\ue}{\mathrm{e}}
\newcommand{\ta}{\ensuremath{t_a}}
\newcommand{\mN}{\ensuremath{\mathcal{N}}} 

\newcommand{\id}{{\mathbbm{1}}}
\newcommand{\tr}{\ensuremath{{\rm{tr}}}}

\begin{document}

\title[Probing Entanglement in Adiabatic Quantum Optimization with Trapped Ions]{Probing Entanglement in Adiabatic Quantum Optimization with Trapped Ions}
\author{Philipp Hauke}
\email{philipp.hauke@uibk.ac.at}
\affiliation{Institute for Theoretical Physics, University of Innsbruck, A-6020
Innsbruck, Austria}
\affiliation{Institute for Quantum Optics and Quantum Information of the Austrian Academy
of Sciences, A-6020 Innsbruck, Austria}
\author{Lars Bonnes}
\email{lars.bonnes@uibk.ac.at}
\affiliation{Institute for Quantum Optics and Quantum Information of the Austrian Academy
of Sciences, A-6020 Innsbruck, Austria}
\author{Markus Heyl}
\email{markus.heyl@uibk.ac.at}
\affiliation{Institute for Theoretical Physics, University of Innsbruck, A-6020
Innsbruck, Austria}
\affiliation{Institute for Quantum Optics and Quantum Information of the Austrian Academy
of Sciences, A-6020 Innsbruck, Austria}
\author{Wolfgang Lechner}
\email{w.lechner@uibk.ac.at}
\affiliation{Institute for Theoretical Physics, University of Innsbruck, A-6020
Innsbruck, Austria}
\affiliation{Institute for Quantum Optics and Quantum Information of the Austrian Academy
of Sciences, A-6020 Innsbruck, Austria}

\begin{abstract}
Adiabatic quantum optimization has been proposed as a route to solve NP-complete problems, with a possible quantum speedup compared to classical algorithms. However, the precise role of quantum effects, such as entanglement, in these optimization protocols is still unclear. We propose a setup of cold trapped ions that allows one to quantitatively characterize, in a controlled experiment, the interplay of entanglement, decoherence, and non-adiabaticity in adiabatic quantum optimization. We show that, in this way, a broad class of NP-complete problems becomes accessible for quantum simulations, including the knapsack problem, number partitioning, and instances of the max-cut problem. Moreover, a general theoretical study reveals correlations of the success probability with entanglement at the end of the protocol. From exact numerical simulations for small systems and linear ramps, however, we find no substantial correlations with the entanglement during the optimization. For the final state, we derive analytically a universal upper bound for the success probability as a function of entanglement, which can be measured in experiment. The proposed trapped-ion setups and the presented study of entanglement address pertinent questions of adiabatic quantum optimization, which may be of general interest across experimental platforms.

\end{abstract}

\maketitle

\section{Introduction}
Owing to an enormous progress in precise preparation, manipulation, and measurement, collections of trapped ions have become a paradigm system for quantum information processing \cite{Wineland1992,Johanning2009,Blatt2012,Schneider2012,Monroe2013}. 
The excellent control over external and internal states allows for the quantum simulation of a broad class of spin models, typically with long-range interactions determined by the structure of the ion crystal  
\cite{Friedenauer2008,Lanyon2011,Islam2012,Khromova2012,Britton2012,Richerme2013b,Richerme2014,Jurcevic2014}. 
Recently, Ising models with disordered long-range interactions have gained interest across research fields because they can represent many NP-complete optimization problems, which have applications ranging from applied computer science to financial markets \cite{Karp1972,Garey1979}. By encoding their cost function into the spin--spin interactions \cite{Lucas2014} these optimization problems could possibly be solved more efficiently by \emph{adiabatic quantum optimization} (AQO), i.e., adiabatic state preparation of a spin-glass ground state \cite{Farhi2001,Santoro2002,Jeremie2002,Amin2008,Hen2014,Heim2014}. 
Procedures in this spirit have been implemented in so-called \textit{quantum annealers} such as the D-Wave device  \cite{Johnson2011,Perdomo-Ortiz2012,Bian2013,Boixo2013,Boixo2014} or nuclear magnetic resonance setups \cite{Steffen2003}. 
However, an unambiguous evidence of the effectiveness and performance of these devices is missing~\cite{Altshuler2010,Boixo2014,Ronnow2014}, both on the fundamental side, i.e., the question whether quantum annealing itself is advantageous over classical algorithms, as well as in view of the unavoidable decoherence present in actual realizations. Although the persistence of entanglement under decoherence has recently been demonstrated in quantum annealing devices \cite{Lanting2014}, the role of entanglement---and in particular the relation between entanglement and the efficiency of obtaining the correct ground state---is far from clear. 

\begin{figure}[t]
\includegraphics[width=\columnwidth]{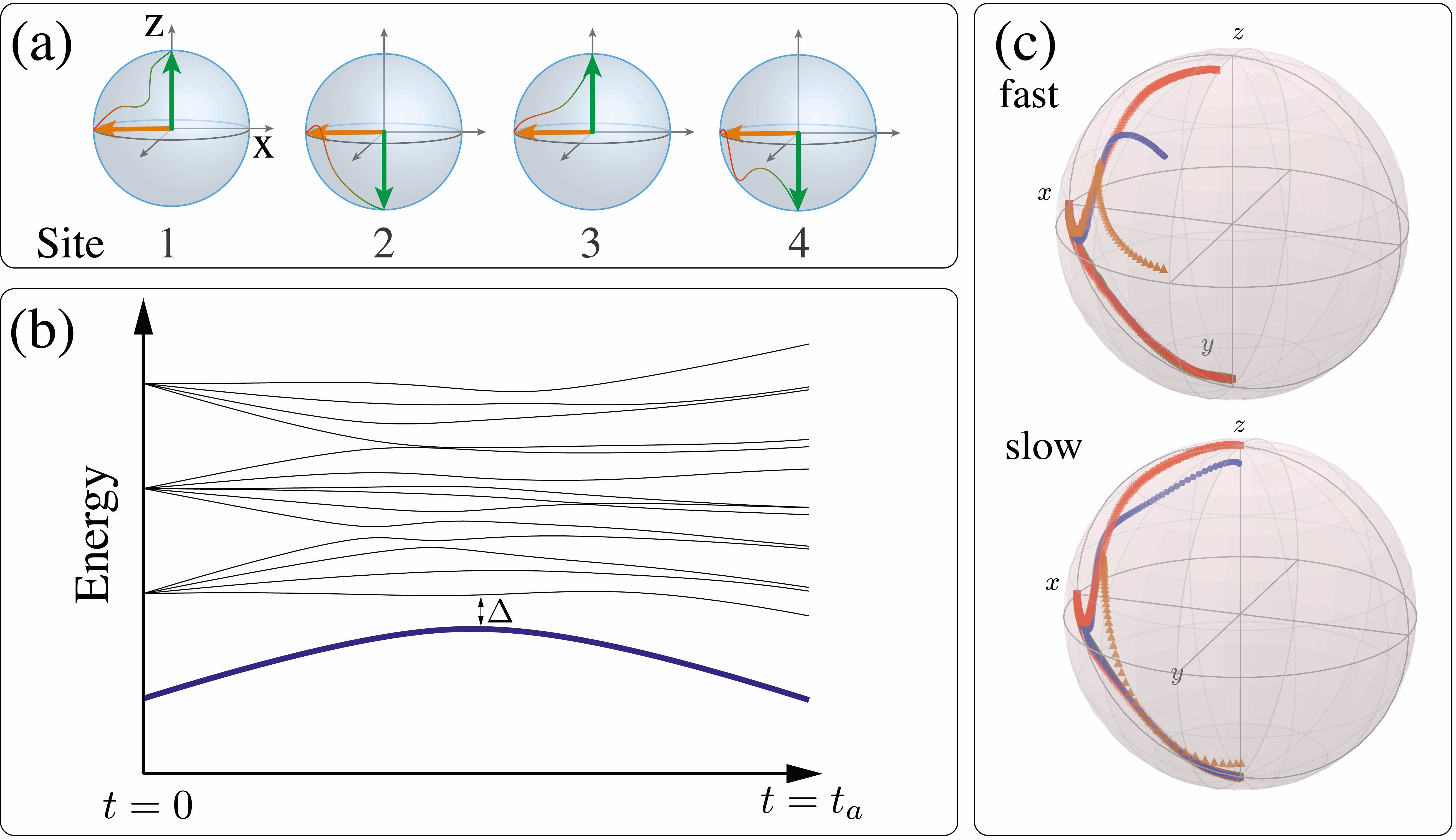}
\caption{(a) In the adiabatic quantum optimization protocol, the system (depicted: four spins represented by their Bloch spheres) is prepared in a known ground state of the simple Hamiltonian $H_{\mathrm{init}}$ (orange arrows pointing to the left). The system is then transferred to the final Hamiltonian $H_{\mathrm{final}}$ with the ground state, given by the green arrows, that is the solution of the optimization problem. (b) If the transformation is performed adiabatically and shielded from decoherence effects, the system remains in the instantaneous ground state (thick blue) protected by the gap $\Delta$. (c) Evolution of the state vectors of four spins (indicated by different colours) on the Bloch sphere for a typical choice of interactions $J_{ij}$. For slow annealing ($\ta=512/J$, bottom), the protocol does reach the final ground state at the poles of the Bloch sphere, while it fails to do so for fast annealing ($\ta=64/J$, top).}
\label{fig:fig1}
\end{figure}

The purpose of this paper is twofold: 
(i) We propose an experimentally feasible implementation based on trapped ions for a variety of famous NP-complete problems. We show how controlled noise can be engineered in such setups. This, in combination with large intrinsic coherence times and precise control, enables a systematic study of how decoherence influences entanglement and efficiency in AQO protocols.  
(ii) We perform a theoretical analysis of the role of entanglement in AQO. We numerically study its interrelation with the success probability, and we analytically show that the entanglement entropy provides an upper bound for the success probability. 

A key question that becomes accessible in the proposed trapped-ion setup is how entanglement, decoherence, and non-adiabaticity on the one side connect to the efficiency on the other side. 
Our findings, obtained from exact numerical calculations, suggest that entanglement in the final state can reveal information about the efficiency of the protocol. We derive a universal upper bound that allows for an efficient estimation of the success probability from the final-state entanglement. Moreover, we find that the maximal entanglement during the sweep is rather uncorrelated with the success probability. Note that the goal of this paper is not to suggest a scalable quantum-annealing device that could compete with system sizes of a D-wave machine \cite{Ronnow2014}, but rather a well-controlled implementation that allows one to study the fundamentals of AQO. 

The basic idea behind AQO is to utilize quantum adiabaticity for solving hard optimization problems that can be encoded in couplings $J_{ij}$ and weights $h_i^z$ of a classical Ising Hamiltonian \cite{Farhi2001,Santoro2002,Jeremie2002,Amin2008} 
\begin{equation}
\label{eq:Hfinal}
H_\mathrm{final} = \sum_{i\neq j} J_{ij} \sigma_i^z \sigma_j^z+\sum_i h_i^z \sigma_i^z\,,
\end{equation} 
where $\sigma_i^\nu$, $i=1\dots N$, denote Pauli matrices. 
The interaction matrix $J_{ij}$ and the magnetic fields $h_i^z$ are chosen such that the ground  state of $H_\mathrm{final}$ is the optimal solution of the original problem.
To arrive at the final ground state, AQO employs an adiabatic sweep starting from a simple to prepare ground state of some $H_{\mathrm{init}}$ (e.g., $H_{\mathrm{init}} = h^x \sum_i \sigma_i^{x}$ with all spins initially polarized along the $x$ direction). The Hamiltonian is then deformed adiabatically such that $H(t=\ta)=H_\mathrm{final}$ after the annealing time $\ta$.
This protocol can be described by the time dependent Hamiltonian  
\begin{equation}
H(t) = A(t) H_\mathrm{init} + [1 - A(t)] H_\mathrm{final}\,, 
\label{eq:HamiltonianAdiabaticPassage}
\end{equation} 
where $A(t)$ is ramped from initially $A(t=0)=1$ to $A(t=\ta)=0$. 
If the ramp is sufficiently slow, according to the adiabatic theorem, the system will remain at all times in the ground state and the state at $t=\ta$ is the solution of the optimization problem. 
This procedure is sketched in Fig.~\ref{fig:fig1}. 
It is in spirit very similar to adiabatic state preparation, where one seeks to reach the ground state of a quantum many-body Hamiltonian via a ramp as Eq.~\eqref{eq:HamiltonianAdiabaticPassage}. First steps in this direction have already been done in a trapped-ion setup \cite{Richerme2013b}. Here, in contrast one seeks to employ quantum fluctuations for solving a classical optimization problem.  

The AQO scheme becomes particularly appealing for complex problems where $H_\mathrm{final}$ is characterized by a high roughness of the free-energy landscape. 
Such problems are equivalent to spin glasses, where classical annealing is known to experience a dramatic slow down at low temperatures because energy barriers become exponentially large \cite{Binder1986}. 
By overcoming the barriers via quantum tunnelling~\cite{Battaglia2006} instead of thermal activation, quantum annealing as defined in Eq.~(\ref{eq:HamiltonianAdiabaticPassage}) offers the prospect of outperforming classical annealing  \cite{Santoro2002}.
However, the speed-up of quantum annealing has been only shown for certain classes of problems \cite{Jeremie2002,Amin2008} and for devices running at finite temperature AQO has been put into question entirely \cite{Katzgraber2014,Ronnow2014}. 
In all of this, the precise role of quantum effects during the sweep is an open question, although the presence of entanglement during quantum annealing at finite temperature has recently been demonstrated in superconducting qubits \cite{Lanting2014}. 

A measure for the efficiency of AQO protocols is the success probability $P$, defined as the overlap between the final state $\ket{\psi(t_a)}$ with the actual ground state $\ket{\psi_{\rm gs}}$ of $H_{\rm final}$. 
Note that this measure can only serve as a benchmark, since it requires a priori knowledge about the solution of the problem. 
In addition, we also consider the \textit{instantaneous} success probability $P(t)$, which monitors the overlap of the state at time $t$ with the desired final ground state, i.e., 
\begin{equation}
P(t)=|\bra{\psi(t)} \psi_{\rm gs} \rangle|^2.
\end{equation}
At the end of the protocol at time $t=\ta$, this quantity coincides with the success probability, $P(\ta)=P$. 

Before presenting the results, we give a short overview of the structure of this paper. In Section~\ref{sec:trappedIons}, we present  feasible implementations of various NP-complete problems in trapped ions and show how noise can be engineered in a controlled manner. In Section~\ref{sec:entanglementObservables}, we introduce relevant entanglement measures that we use in this work. Analytical results on the relation between entanglement after the sweep and the success probability  are presented in Section~\ref{sec:upperBound}. Numerical results for closed and open systems are discussed for a model problem, the Coulomb glass, in Section~\ref{sec:numericalResults}. We close by presenting our conclusions in Section~\ref{sec:conclusions}.

\section{Quantum Adiabatic Optimization with Trapped Ions\label{sec:trappedIons}}

In this section, we present how current trapped-ion technology \cite{Friedenauer2008,Lanyon2011,Islam2012,Khromova2012,Britton2012,Richerme2013b,Richerme2014,Jurcevic2014} can be extended to implement several well-known NP-complete problems \cite{Karp1972,Garey1979}. The main challenge is to encode these problems in $H_\mathrm{final}$ [Eq.~(\ref{eq:Hfinal})], which requires a certain degree of programmability of interactions $J_{ij}$ and fields $h_i^z$. Below, we present several models where a local variation of laser power is sufficient to obtain this programmability---as opposed to a much more difficult programming by, e.g., an extensive number of laser frequencies \cite{Korenblit2012} or specially designed trapping potentials~\cite{Chiaverini2008,Schmied2009b,Zippilli2014,Wilson2014}. We also describe how engineered noise can be generated in order to study the interplay between decoherence, non-adiabaticity, and entanglement in the well-controlled architecture provided by trapped ions. 

\subsection{Ion Hamiltonian\label{sec:IonHamiltonian}}

We consider a chain of ions confined in a linear Paul trap. The spin-1/2 Ising variables appearing in $H_{\mathrm{final}}$ can be encoded by restricting the internal atomic dynamics to only two hyperfine sublevels. 
Single-qubit rotations allow for a preparation with high fidelity of any desired initial product state \cite{Jurcevic2014}, including the completely polarized ground state of $H_{\rm init}$ \cite{Friedenauer2008,Lanyon2011,Islam2012,Britton2012,Richerme2013b,Richerme2014,Jurcevic2014}.  
Similarly, the projection of the spin-1/2 variable on any coordinate direction can be measured to high precision by appropriate single-qubit rotations followed by stimulated fluorescence measurements, allowing the reconstruction of arbitrary correlation functions \cite{Lanyon2011,Jurcevic2014}. 
The synthetic magnetic fields $h_i^z$ and $h^x$ appearing in $H_{\mathrm{final}}$ and $H_{\mathrm{init}}$ can also be simulated in a straightforward fashion via AC-Stark shifts or by resonantly driving the transition between the two sublevels \cite{Porras2004a}. 

Ising interactions can be simulated by coupling the ions via laser \cite{Cirac1995,Sorensen1999a,Porras2004a} or microwave radiation \cite{Mintert2001} to the collective vibrational modes. In the limit of off-resonant radiation, one can disentangle the collective vibrations and the atomic Ising variables via a canonical transformation and obtain an effective Hamiltonian for the Ising variables alone. For example, when employing a M{\o}lmer--S{\o}rensen scheme \cite{Sorensen1999a}, the Ising spins are subject to the effective interactions (see Appendix~\ref{sec:AppendixIsingInteractionsInIons}) 
\begin{equation}
\label{eq:HIsingIons}
J_{ij} = - \hbar\frac{|\Omega_i| |\Omega_j|}{4}\sum_q \frac{\eta_{iq} \eta_{jq}}{\delta_q}\,,
\end{equation}
where $\eta_{iq}$ is Lamb--Dicke parameter including the amplitude of vibrational mode $q$ at ion $i$, $\delta_q$ is the detuning of the laser relative to the vibrational mode $q$, and $|\Omega_i|$ is the absolute value of the laser's Rabi frequency at ion $i$. 
For the derivation of Eq.~\eqref{eq:HIsingIons}, we assumed the off-resonance condition $\eta_{iq}|\Omega_i|\ll \delta_q$. The addition of a transverse field, required for our choice of $H_{\mathrm{init}}$, induces additional couplings to the phonon modes in higher order, which, however, can be neglected as long as $h_i^\nu\ll \hbar\delta_q$, $\nu=x,z$.
The precise form of the interactions depends on the coupling scheme used, but the qualitative behavior is similar for other schemes \cite{Porras2004a,Mintert2001}. 

As shown, e.g., in Refs.~\cite{Britton2012,Islam2012,Jurcevic2014}, when transmitted via the transverse phonon modes the Ising interactions follow an approximate power law as a function of the distance, $J_{ij} \sim 1 / |i-j|^\alpha$, with a decay coefficient $0\leq \alpha\leq 3$ that can be adjusted using the detuning $\delta_q$ (where larger detuning entails larger $\alpha$).  
When employing the longitudinal phonon modes, the large spacing of vibrational frequencies restricts the range to $\alpha\approx 0$. 
By expressing the local Rabi frequencies as multiplies of a reference frequency $|\Omega|$, i.e., $|\Omega_i|=F_i |\Omega|$, we can thus write  
\begin{equation}
\label{eq:JijIons}
J_{ij} = J \frac{F_i\, F_j} { |i-j|^\alpha }\,,
\end{equation} 
with $J$ setting the overall energy scale. 
Note that the power-law decay is exact only in the limits $\alpha=0$ and $\alpha=3$. As shown recently in Ref.~\cite{Nevado2014}, the interactions in the intermediate regimes are better described by an exponential tail added to a dipolar decay. Additional modifications may arise from an uneven spacing of ions as is usual in linear Paul traps or from inhomogeneous laser profiles. For small chains, a general power-law behaviour is, however, a good approximation of the system dynamics \cite{Jurcevic2014}.

Until now, we considered a coupling to a single set of vibrational modes. However, richer models can be quantum simulated by employing the fact that in a linear Paul trap the three orthogonal directions of vibration decouple \cite{Porras2004a}. In this case, by using different lasers to couple to each orthogonal set of modes, it is possible to generate three independent Ising interaction terms, which we label $\ell=1,2,3$. 
Additionally, one may induce synthetic fields $h_i^z$ in an ion-dependent fashion by, e.g., resonantly driving the qubit transitions with beams derived from an acousto-optic deflector \cite{Jurcevic2014}.  
It is thus possible to realise in trapped ions---using available technology---the following Ising Hamiltonian:  
\begin{equation}
\label{eq:Htrappedions}
H_{\mathrm{final}}=\sum_{\ell=1,2,3} J \sum_{i\neq j}\frac{F_i^{(\ell)} F_j^{(\ell)}}{\left|i-j\right|^{\alpha^{(\ell)}}} \sigma_i^z\sigma_j^z + \sum_{i}h_i^z \sigma_i^z \,. 
\end{equation}

\subsection{NP-complete models realizable with available trapped-ion technology}

\begin{table*}
\begin{tabular}{ l l }
\toprule
Optimization problem \qquad\quad 	& 	Parameters for Hamiltonian Eq.~\eqref{eq:Htrappedions}	\tabularnewline
\cmidrule{1-2}
{$\begin{array}{l}
\vspace*{-0.25cm}\\
\mathrm{constrained}\\
\mathrm{Coulomb}\\
\mathrm{glass}
\end{array}$}			
					& 	
						$\begin{array}{l}
						\vspace*{-0.3cm}\\
						\alpha^{(1)}=1\,,\,\,\quad F_i^{(1)}\equiv 1\\
						\alpha^{(2)}=0\,,\,\,\quad F_i^{(2)}\equiv F^{(2)}\gg 1 \\
						h_i^z \in[-\epsilon_z,\epsilon_z]\,\, \mathrm{random}\,;\,\, \epsilon_z=\mathcal{O}(J)
						\end{array}$ 
						\vspace*{0.5cm}
											\tabularnewline
$\begin{array}{l}
\vspace*{0.05cm}
\mathrm{Number}\\
\mathrm{partitioning}
\end{array}$
					& 
						$\begin{array}{l}
						\vspace*{0.05cm}
						\alpha^{(1)}=0\,,\,\,\quad F_i^{(1)}= n_i\\
						h_i^z\equiv 0 
						\end{array}$ 
						\vspace*{0.1cm}
											\tabularnewline								
$\begin{array}{l}
\vspace*{0.0cm}\\
\mathrm{Integer}\\
\mathrm{knapsack}
\end{array}$
\vspace*{-0.cm}
					&
						\vspace*{-0.cm}
						$\begin{array}{l}
						\vspace*{-0.cm}
						\alpha^{(1)}=0\,,\,\,\quad F_i^{(1)}= 
							\left\{\begin{array}{l}
							i \,, \quad\,\,\,\, i=1\dots C \\
							w_i \,, \quad i=C+1\dots N
							\end{array}\right.
						\end{array}$
						\vspace*{0.cm}
											\tabularnewline
					&
						$\begin{array}{l}
						\vspace*{0.2cm}
						\alpha^{(2)}=0\,,\,\,\quad F_i^{(2)}= 
							\left\{\begin{array}{l}
								1 \,, \quad\hspace*{0.175cm} i=1\dots C \\
								0 \,, \quad\hspace*{0.175cm} i=C+1\dots N
							\end{array}\right.
						\\
						h_i^z = J
							\left\{\begin{array}{l}
								2(C - 2) + i \left[C(C+1)  - 2 \sum_{j=C+1}^N w_j\right] \,, \,\, i=1\dots C \\
								2 c_i\, J^\prime / J + w_i \left[C(C+1)  - 2\sum_{j=C+1}^N w_{j}\right] , \,\, i=C+1\dots N
							\end{array}\right.
						\\
						\end{array}$
											\tabularnewline
\bottomrule
\end{tabular}
\caption{
{ 
NP-complete problems encodable via Hamiltonian Eq.~\eqref{eq:Htrappedions}.} 
\textmd{
(a) \emph{Coulomb glass problem}, which can be mapped to a class of opti-cut problems \cite{Anjos2013}. 
(b) \emph{Number-partitioning problem}: Given a set $\mathcal{N}$ of positive integers $n_i\in \mathbb N$, $i=1,\dots,N$, find a subset $\mathcal{R}\subset \mathcal{N}$ such that the sum over all numbers in $\mathcal{N}$ and the remainder $\mathcal{N} \backslash \mathcal{R}$ is equal, i.e., $\sum_{i\in \mathcal{R}} n_i =  \sum_{i\in\mathcal{N} \backslash \mathcal{R}} n_i$.  
(c) \emph{Integer knapsack problem}: Given a set $\mathcal{M}$ of objects $k=1,\dots,M$, with associated weights $w_k\in \mathbb N$ and values $c_k\in \mathbb R$, and a container (a ``knapsack'') with maximal capacity $C\in \mathbb N$, how full can the knapsack be filled without making it overflow? That is, find the subset $\mathcal{R}\subset \mathcal{M}$ that maximizes the total cost $c_{\mathrm{tot}} \equiv \sum_{k\in \mathcal{R}} c_k$ (or the total weight $w_{\mathrm{tot}}\equiv \sum_{k\in \mathcal{R}} w_k$), subject to the constraint $w_{\mathrm{tot}}\leq C$. 
To encode this problem, the first $C$ spins represent the obtained integer filling (only one of these spins is in state $+1$), and the remaining $M=N-C$ spins encode if an item is included in the knapsack ($-1$) or not ($+1$). The cost function is only encoded in the local field term $2 c_i\, J^\prime $, while the rest of the terms serve to enforce the constraint $w_{\mathrm{tot}}\leq C$ on the container, for which we require $0<J^\prime \max_k(c_k)<J$. 
More details for the encoding of the knapsack problem in Ising interactions can be found in Ref.~\cite{Lucas2014}. It may be advantageous to restrict the weights to an interval, since this can increase the hardness of the instances \cite{Pisinger2005} while simultaneously reducing the spread of interaction strengths $J$}.
\label{tab:NPcompleteProblems}}
\end{table*}

Experiments thus far---aiming at mimicking translationally invariant many-body models---made great efforts to engineer systems as homogeneous as possible, i.e., $F_i^{(\ell)}=1$ and $h_i^z=h^z$, $\forall\, i$ \cite{Lanyon2011,Islam2012,Britton2012,Richerme2014,Jurcevic2014}. 
For the same reason, few theoretical works have considered a site-dependent tunability of couplings \cite{Bermudez2011,Bermudez2012b,Hauke2013b,Zippilli2014} or fields \cite{Hauke2014}. 
In contrast, our purpose of engineering NP-complete problems in Hamiltonian Eq.~(\ref{eq:Hfinal}) requires a certain degree of programmability of the interaction matrix elements $J_{ij}$ and fields $h_i^z$ (see, e.g., Ref.~\cite{Lucas2014}). 
In Table~\ref{tab:NPcompleteProblems}, we present several NP-complete problems that can be implemented with interactions of the form given in Eqs.~\eqref{eq:JijIons} and~\eqref{eq:Htrappedions}, i.e., without requiring a full programmability of interactions. These problems thus become amenable to current trapped-ion technology by adjusting the local laser intensities $|\Omega_i|^2$. 

The NP-complete problems implementable in this way include the \emph{number-partitioning problem}, the \emph{integer knapsack problem} \cite{Lucas2014}, and the \emph{Coulomb glass problem}, which---as shown in Ref.~\cite{Anjos2013}---can be mapped to a class of \emph{opti-cut problems}. 
We summarize the description of the necessary Hamiltonian parameters in Table~\ref{tab:NPcompleteProblems}. In the numerical studies presented in Sec.~\ref{sec:numericalResults}, we focus on the example of the Coulomb glass problem, since it is closest to current experiments \cite{Richerme2014,Jurcevic2014}. The Coulomb glass \cite{Ortunyo2004} contains a random magnetic field $h_i^z  \in [-\epsilon_z,\epsilon_z]$ and Ising interactions that are homogeneous and long-ranged, $\alpha^{(1)}=1$ and $F_i^{(1)}\equiv 1$. Usually, one imposes the constraint of vanishing total magnetization, which can be realized by mean-field-like interactions ($\alpha^{(2)}=0$) with $F_i^{(2)}\equiv F^{(2)}\gg 1$.  
Defining the strength of the constraint as $V\equiv J (F^{(2)})^2 \gg J$, the total Hamiltonian is then 
\begin{equation}
\label{eq:CoulombGlass}
H_\mathrm{final}=J\sum_{i\neq j}\frac{\sigma_i^z\sigma_j^z}{|i-j|}+ \sum_i h_i^z\sigma_i^z + V\sum_{i\neq j}\sigma_i^z\sigma_j^z\,, 
\end{equation}
with $J,V>0$.
The large frustration of the Ising interactions counteracts ordering tendencies and, for $\epsilon_z=\mathcal{O}(J)$, leads to a strong competition between the interactions and the random local fields. This competition signs responsible for its computational complexity. Currently, it is not clear whether the Coulomb glass exhibits a spin-glass transition at nonzero temperature. This, however, does not necessarily imply a reduced complexity of the ground state of the model. Certain random Ising models on Chimera graphs, such as implemented in the D-Wave device, for example, do not show spin-glass transitions at nonzero temperature, but finding the ground state of the model is still NP-hard. Additionally, it may be that a non-perfect power-law decay, for example the more realistic dipolar plus exponential shape \cite{Nevado2014} affects the complexity of the model.

\subsection{Noise engineering}

A fundamental question in AQO is how the optimization of problems such as given in Table~\ref{tab:NPcompleteProblems} performs in the presence of decoherence. This question is of particular importance for current implementations of AQO devices such as the D-Wave machine where decoherence rates are large compared to the annealing times.
Current trapped-ion setups are able to suppress noise to very low values, currently with time scales around $10/J$ \cite{Jurcevic2014}. This opens the possibility to systematically study the performance of AQO under varying degrees of purposefully engineered decoherence. A scheme to engineer different types of classical noise in a controlled fashion has recently been discussed and implemented in Ref.~\cite{Soare2014}. 

Here, we describe how classical white noise can be engineered by a conceptually simple extension of the steps leading up to Eq.~\eqref{eq:HIsingIons}. Detailed derivations can be found in the Appendix~\ref{sec:AppendixNoiseInIons}. Here, we sketch the main idea, which is based on the addition of randomly fluctuating synthetic magnetic fields. 
We assume that the rate of these fluctuations is much larger than any other parameter in the Hamiltonian, so that we can treat the annealing parameter $A$ as quasi-constant. The total Hamiltonian then takes the form
\eq{
H = A H_{\mathrm{init}} + (1-A) H_{\mathrm{final}} + \sum_i\left[ \xi_i^x(t)\sigma_i^x + \xi_i^z(t)\sigma_i^z  \right]\,,
} 
where we choose the noise in two spin directions $\nu=x,z$, and we take $\xi_i^\nu(t)$, to be independent Ornstein--Uhlenbeck processes, $\llangle\, \xi_i^\nu(t) \xi_j^\mu(0) \,\rrangle=\hbar\Gamma \delta_{\nu,\mu} \delta_{i,j} \hbar b \ue^{-b |t|}$. Here, $\llangle\, \cdot \,\rrangle$ denotes the average over noise realizations, and for simplicity we take the rate $\Gamma$ and the bandwidth $b$ to be independent of $\nu=x,z$.  
The fluctuating fields can be realized using the same experimental techniques as for their quasi-static counterparts described at the beginning of Sec.~\ref{sec:IonHamiltonian}. The additional requirement of fast fluctuations that are uncorrelated between different sites could be solved via a simultaneous addressing of the individual ions using several laser beams derived from an acousto-optic deflector \cite{Jurcevic2014}. 

If the bandwidth $b$ of the fluctuations is much larger than the internal energy scales of the ions and the coupling to the phonon modes, $b\gg h_i^\nu/\hbar, \eta_{iq} |\Omega_i|$, we can average over the noise and obtain a master equation describing classical white-noise dephasing, 
\begin{equation}
\label{eq:ME_general}
\frac{\partial \rho}{\partial t} = -\frac{i}{\hbar} [H,\rho] + \Gamma \sum_i D[\sigma^{(x)}_i] + \Gamma \sum_i D[\sigma^{(z)}_i]\,,
\end{equation}
where $D[X] = 2 X\rho X^\dagger - \rho X^\dagger X - X^\dagger X\rho$ is the Lindblad super-operator. 
In the derivation of \Eref{eq:ME_general}, we neglected additional couplings to the phonon modes that appear in higher order and can be dropped provided 
 $b \gg \eta_{iq}|\Omega_i|,h_i^{\nu}/\hbar$. 
These conditions complement the requirements for the derivation of $A H_{\mathrm{init}} + (1-A) H_{\mathrm{final}}$, which read $\delta_q\gg \eta_{iq}|\Omega_i|, h_i^\nu/\hbar$. In the Appendix~\ref{sec:AppendixNoiseInIons}, we discuss that they can all be fulfilled for realistic experimental parameters. 

Hence, adding fluctuating fields with bandwidth much larger than other relevant energy scales allows to systematically engineer white noise in trapped ions. By inducing such fluctuations on the longitudinal or transversal fields, it is possible to study different types of Lindblad super-operators.

\section{Entanglement Observables\label{sec:entanglementObservables}}

AQO aims at solving classical problems using a quantum device, prompting the fundamental question of the role and importance of quantum effects within such a computational scheme. One particular feature of quantum as opposed to classical systems is entanglement, a theoretical study of which is the second central aim of this work. For that purpose, we use different quantifiers for quantum many-body correlations: the entanglement entropy, the logarithmic negativity, an optimal local entanglement witness, and the Fisher information. All of these quantifiers are sensitive towards different forms of entanglement and thus allow to address different aspects of entanglement. 

In a perfect adiabatic protocol, entanglement vanishes in the initial and the final state, but it is finite during the sweep. The intermediate entanglement is due to avoided crossings between the ground state and excited states. In an optimal sweep, the final state shows no entanglement at all, as the true ground state is that of a classical Ising model. Therefore, entanglement in the final state is an indication for imperfections during the sweep. This mechanism of build-up and release of entanglement during an adiabatic optimization is the focus of the remainder of this work. 

A full study of the many-body entanglement, which would require a full state tomography, is computationally and experimentally not feasible for all but very small systems. However, one can gain insight into different aspects of the problem from the combination of several entanglement measures and witnesses. In this chapter, we will give an overview of the various entanglement observables that we use. We also introduce a novel witness that can be efficiently measured in experiment. 
\\

\textit{Entanglement Entropy.---} A convenient, well-established measure for entanglement of closed systems is the entanglement entropy, which is defined between two subsystems $A$ and $B$. The entanglement entropy is the von Neumann entropy of the reduced density matrix $\rho_A = \tr_B \rho$ of $A$ obtained by tracing out the complement $B$, 
\begin{equation}
	S_A = -\tr\left[ \rho_A \log(\rho_A)\right]\,.
	\label{eq:vNEntropy}
\end{equation}
In this work, we study the maximum of the entanglement entropy considering all possible pairs of particles as subsystem $A$ and the remaining $N-2$ particles as subsystem $B$. This pair-block entanglement entropy maximized over all pairs is denoted as $S_2^{\textrm{max}}(t)$. The largest pair-block entanglement entropy reached during the sweep is $S_2^{\textrm{max}}=\max_t \left[ S_2^{\textrm{max}}(t) \right]$.

\textit{Logarithmic Negativity.---} A convenient mixed-state entanglement measure is the logarithmic negativity between two parts $C$ and $D$ of a larger subsystem $A$ ($A= C \cup D$). It is defined as \cite{Vidal02,Plenio2005} 
\begin{equation}
	E_\mN = \log || \rho_A^{T_C}||\,,
\end{equation}
where $||X|| = \tr \sqrt{X^\dag X}$ is the trace norm. $\rho_A^{T_C}$ is the partial transpose of the reduced density matrix $\rho_A$ with respect to subpart $C$, given by $\langle d c | \rho_A^{T_C} | d' c' \rangle = \langle d c'| \rho_A | d' c \rangle$ where the $c$'s and $d$'s form a basis for $C$ and $D$, respectively. 
In the following, we consider the logarithmic negativity between subsystems $C$ and $D$ consisting of a single site each, and maximize again over all combinations. 
\\

\textit{Fisher Information.---} While entropy and negativity are measures for the entanglement between two subblocks of the system, we are also interested in multi-particle entanglement. This can be measured with the Fisher information $F_Q$, which not only quantifies entanglement, but also provides a lower bound on the number of particles that are entangled \cite{Hyllus2012,Smerzi2009}: if a state is $k$-producible---meaning that it can be written as a product of states involving less than or equal to $k$ particles each---the Fisher information satisfies  
\begin{equation}
F_Q \leq \lfloor  \frac{N}{k} \rfloor k^2 + \left(N - \lfloor  \frac{N}{k} \rfloor k \right)^2 \,.
\end{equation}
Here, $N$ is the total number of spins and $\lfloor  \cdot  \rfloor$ denotes the \textit{floor} operation. If $F_Q > N$, this inequality implies entanglement between an increasing number of particles with increasing $F_Q$, where $F_Q$ close to the maximal value of $N^2$ corresponds to $N$-particle entanglement. In quantum annealing devices, the multi-particle nature of entanglement may be a crucial aspect, since in a spin glass, for example, the lowest-energy configurations are typically vastly different. Thus, many spin flips should be necessary to tunnel from one local energy minimum to the next. If one assumes that all of these spins have to be coherent during the tunnelling process, multi-particle entanglement may become the  critical phenomenon for the efficiency of AQO protocols. 
\\

\textit{Optimal Local Entanglement Witness.---} While the above entanglement observables are powerful tools to theoretically understand the nature of entanglement, they require elaborate measurements in experiment. Therefore, it is desirable to find entanglement witnesses that can be deduced from simple observables. 

Here, we introduce an \emph{optimal local entanglement witness} $W$ that allows one to measure entanglement in closed as well as open systems. It can be obtained from two-point spin correlations,  
\begin{equation}
	W = \max\left\{0,- \min_{\vec f} \left[ W_{\vec f} \right]\ \right\},
\end{equation}
with
\begin{equation}
	W_{\vec f} = \frac{1}{2}\braket{\hat{C}_{\vec f}}-\id, \,\,\hat C_{\vec f} =  \sum_{i,j=1}^N \sum_{\nu=x,y,z} f_i^{\star}f_j \hat{\sigma}_{i}^{\nu}\hat{\sigma}_{j}^{\nu}\,,
\end{equation}
where $f_i\in \mathbb{C}$ is subject to the normalization $\sum_i |f_i|^2=1$.  This constitutes a generalization of the witnesses introduced in Refs.~\cite{Krammer2009,Cramer2011,De-Chiara2011a,Hauke2012g}. Going beyond Ref.~\cite{Hauke2012g}, we analyze here the ``optimal'' witness, obtained by minimizing $W_{\vec f}$ over all $\vec f$. 
In practical terms, this minimization is extremely simple: by introducing a Lagrange multiplier for the condition of normalization $\sum_i |f_i|^2=1$, it becomes equivalent to finding the smallest eigenvalue of the correlation matrix $\mathcal{C}_{ij}=\sum_\nu\braket{\hat{\sigma}_{i}^{\nu}\hat{\sigma}_{j}^{\nu}}$. 
This procedure can be done via post-processing of the measured two-point correlation functions, which are simple observables in trapped-ion experiments \cite{Friedenauer2008,Lanyon2011,Islam2012,Britton2012,Khromova2012,Richerme2013b,Richerme2014,Jurcevic2014}. Thus, it is straightforward to determine the optimized entanglement witness experimentally. Since no cross-correlations of the form $\braket{\hat{\sigma}_{i}^{x}\hat{\sigma}_{j}^{y}}$ are required, it needs only three series of measurements for the $x$, $y$, and $z$ correlations respectively, and is therefore easily scalable.

The above entanglement witness gives a lower bound on the best separable approximation (BSA)~\cite{Cramer2011}, an entanglement measure that is applicable also to mixed states. The ease of measuring this witness, both in theory and experiments, may make it an important entanglement quantifier, especially for large systems.

\section{Upper Bound on the Efficiency from Entanglement}
\label{sec:upperBound}

At the end of an ideal adiabatic protocol, the system arrives at the desired solution of the classical problem with vanishing entanglement. In presence of imperfections such as a non-adiabatic transfer of state population to higher energies, however, the final state will contain contributions from several eigenstates of $H_{\mathrm{final}}$ and in consequence a nonzero entanglement. It is the aim of this section is to show that it is possible to quantify this connection between entanglement and success probability in terms of the following bound
\begin{equation}
	S_l^{\mathrm{end}} \leq -P\log(P) - (1-P) \log(1-P) + (1-P) \log(2^l-1),
\label{eq:upperBound}
\end{equation}
valid for $P>1/e$ with $e$ Euler's number. Here, $S_l^\mathrm{end}$ denotes the final entanglement entropy of an arbitrary subsystem containing $l$ spins. Remarkably, although $S^\mathrm{end}_l$ and $P$ depend on the microscopic details of the system, the bound itself is independent of any microscopic parameters and as such completely universal. Below, we will show numerical data demonstrating that in most cases there is at least one subblock of $l=2$ spins where the bound is closely approached as long as the sweep is sufficiently adiabatic.

We now prove the bound Eq.~(\ref{eq:upperBound}). Let us denote the desired solution of the ideal problem by the spin configuration
\begin{equation}
	|s^\ast\rangle = |s_A^\ast s_B^\ast\rangle\,,
\label{eq:upperBoundState}
\end{equation}
where $|s_{A/B}^\ast\rangle$ denotes the spin configurations in the respective subsystems $A$ and $B$. This product form of the desired solution is always possible because the ground state $|s^\ast\rangle$ of the classical Hamiltonian $H_{\mathrm{final}}$ is a separable state.

In a non-adiabatic sweep, the final reduced density matrix $\rho_A$ of subsystem $A$ will contain diagonal as well as off-diagonal elements. Let us denote by $\rho_A^d$ the diagonal part of the original density matrix $\rho_A$, i.e., where all off-diagonal elements in the $\sigma^z$-basis are set to zero. The associated so-called diagonal entropy $S_A^d$ satisfies the inequality~\cite{Polkovnikov2011uq}
\begin{equation}
	\label{eq:InequalitySASd}
	S_A \leq S_A^d = -\mathrm{tr} [\rho_A^d \log(\rho_A^d)] = -\sum_{\mu=1}^{d_A} p_\mu \log(p_\mu)\,.
\end{equation}
Here, $p_\mu$ denote the eigenvalues of $\rho_A^d$ and $d_A=2^l$ is the size of the Hilbert space of the subblock $A$ consisting of $l$ spins. The magnitude of the diagonal elements can be estimated in the following way. Consider the matrix element $p_1$ defined as 
\begin{equation}
	\label{eq:p1}
	p_1 = \langle s_A^\ast | \rho_A^d |s_A^\ast \rangle = \sum_\nu \langle s_A^\ast \nu | \rho | s_A^\ast \nu \rangle \geq  P\,. 
\end{equation}
The matrix element $p_1$ is the probability to be locally in the ground state, i.e., all spins of the block $A$ are oriented in the direction of the desired solution, Eq.~(\ref{eq:upperBoundState}), and $\nu$ sums over all spin configurations of the complement $B$. The inequality in the above Eq.~\eqref{eq:p1} follows because all diagonal elements of the density matrix $\rho$ are nonnegative. When $P\geq 1/e$, i.e., for not too small success probabilities, Eq.~\eqref{eq:p1} gives $-p_1 \log(p_1) \leq -P \log(P)$, and inequality \eqref{eq:InequalitySASd} becomes 
\begin{equation}
	S_A \leq -P \log(P) - \sum_{\mu=2}^{d_A} p_\mu \log(p_\mu).
\end{equation}
The contribution to the diagonal entropy from the states other than the local ground state assumes its maximum for the case of equipartitioned probabilities, $p_\mu = C/(d_A-1)$, with $\mu\geq 2$ and $C$ a constant. The normalization of the density matrix imposes the constraint $\sum_{\mu=1}^{d_A} p_\mu=1 = p_1+C$. Using $p_1\geq P$, this gives $C\leq 1-P$, which implies
\begin{equation}
	-C \log(C) \leq -(1-P) \log(1-P).
\end{equation}
For a subblock $A$ consisting of $l$ spins, this directly gives the desired upper bound in Eq.~(\ref{eq:upperBound}) by denoting $S_A$ as $S_l^\mathrm{end}$.

An important application of the bound is an estimation of the success probability from entropy measurements with polynomial effort in system size. A direct measurement of the success probability requires knowledge of the desired solution, the exponentially hard problem we set out to solve in the first place. However, the entanglement entropy of a subblock of small size, $l=2$ say, only requires mapping out the reduced density matrix of $l$ spins. This can be performed efficiently even though the total number of spins $N$ might be very large. The maximization over all sets of fixed size $l=2$ leads to an additional increase of measurement resources, but it remains polynomial because the total number of partitions into subblocks of e.g. size $l=2$ is $N(N-1)/2$. 

\begin{figure}[t]
\centering
\includegraphics[width=\columnwidth]{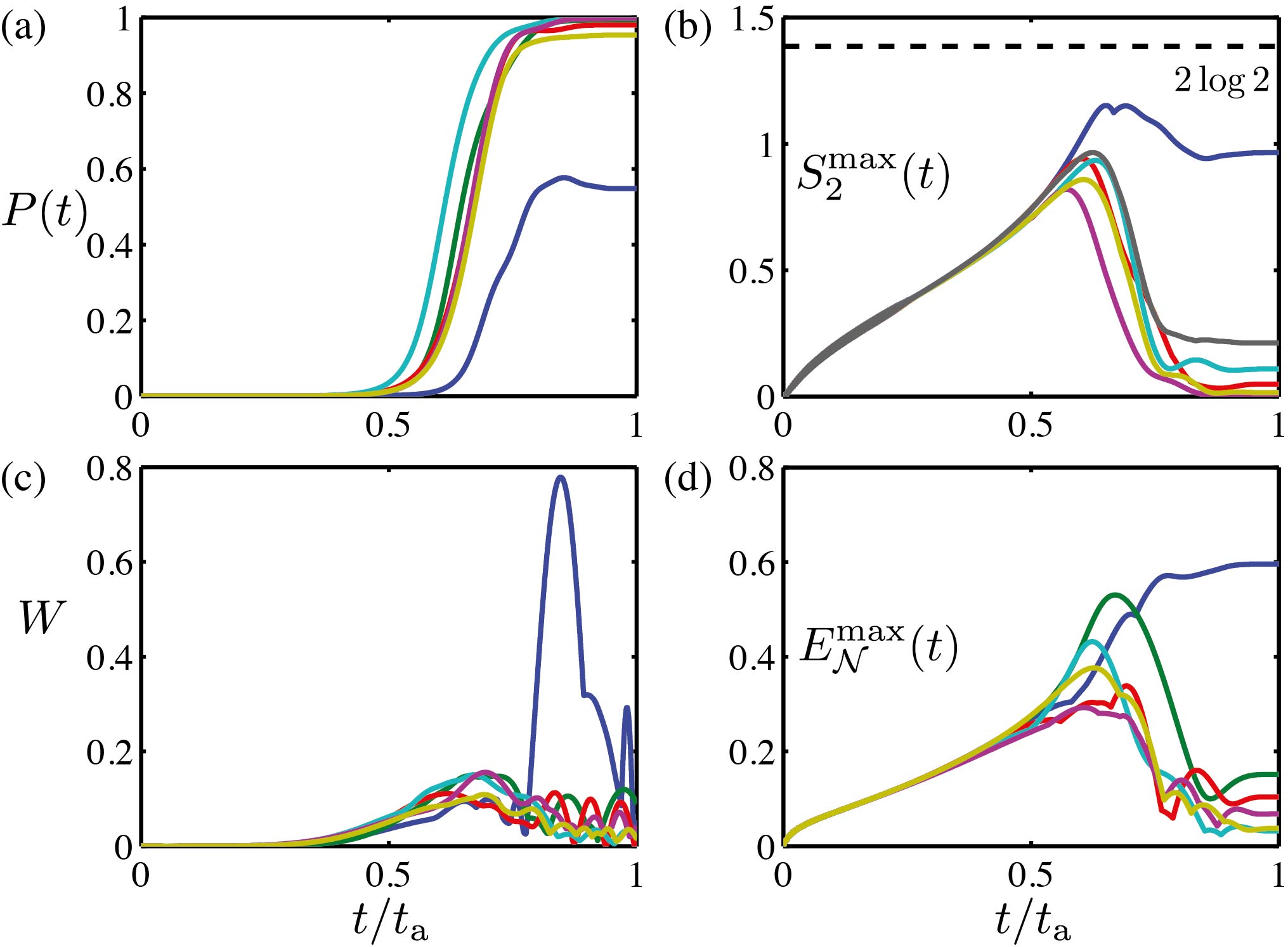}
\caption{
Success probability (a) and entanglement dynamics (b-d) during the adiabatic quantum optimization protocol, for a few instances (shown in different color) of randomly drawn magnetic fields $h_i^z\in [-\epsilon_z,\epsilon_z]$, with $\epsilon_z/J=1/2$. System size is $N=16$, for which the chosen annealing time $\ta=500/J$ is a moderate value. 
(a) The instantaneous success probability, i.e., the overlap with the solution of the optimization problem, increases strongly in the vicinity of the equilibrium magnetic phase transition of the homogeneous system. 
(b) Dynamics of the entanglement entropy $S_2^{\rm max}(t)$ of two spins with the rest,
(c) the optimal local entanglement witness $W$, and (d) the logarithmic negativity $E_{\mN}^\mathrm{max}(t)$ between pairs of spins. Both $S_2^{\rm max}(t)$ and $E_{\mN}^\mathrm{max}(t) $ have been obtained by maximization over all possible permutations of two sites on the chain. Easy instances [large $P(t_a)$] are able to reduce the entanglement obtained around the equilibrium phase transition, while hard instances (blue) remain strongly entangled even towards the end of the protocol. 
\label{fig:entanglementDynamics}
}
\end{figure}

\section{Numerical Results\label{sec:numericalResults}}

We now study numerically the interrelation between entanglement, success probability, and decoherence in the Coulomb-glass problem, Eq.~\eqref{eq:CoulombGlass}. In particular, we show that the bound Eq.~\eqref{eq:upperBound} is a useful measure for the success probability.   
As initial state of the AQO protocol, we choose the ground state of $H_{\mathrm{init}} = h^x \sum_i \sigma_i^{x}$, a classical product state, and we employ a linear ramp, i.e., $A(t)=1-t/\ta$. Note, that possible optimized non-linear ramps may improve the performance of the AQO protocol \cite{Jeremie2002,mandra2014}. We use $\epsilon_z/J=1/2$ and $V/J=10$ as system parameters. 
Before turning to open systems described by the master equation~\eqref{eq:ME_general}, it is instructive to understand the decoherence-free case, which is the subject of the following section. 

\subsection{Closed Quantum Systems}

\begin{figure}[t]
\centering
\includegraphics[width=\columnwidth]{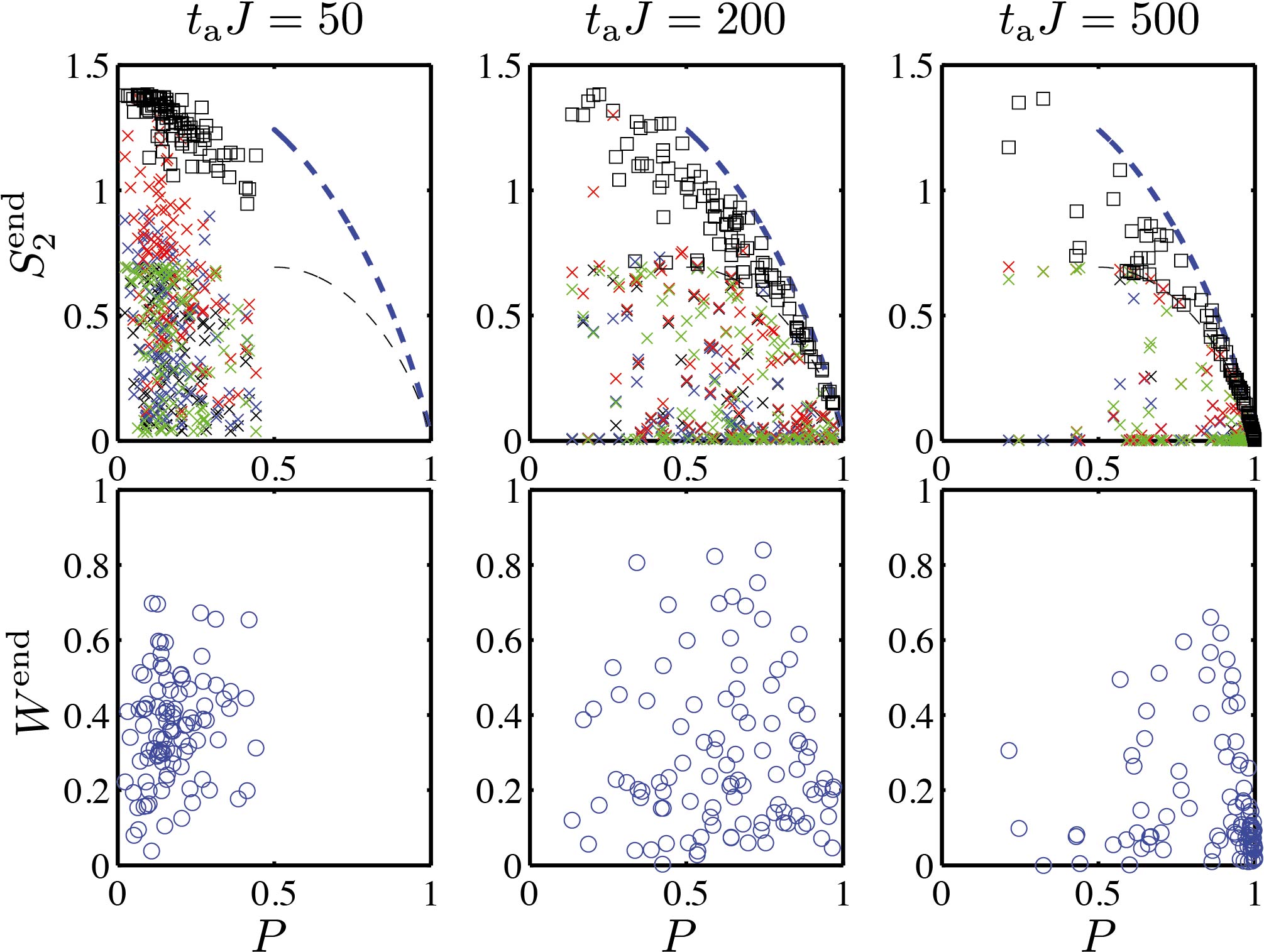}
\caption{
Success probability as a function of final entanglement for different annealing times $\ta$, in a closed system (parameters as in Fig.~\ref{fig:entanglementDynamics}).
{\bf Top:} Final entanglement entropy $S^\mathrm{end}_2$ of exemplary subsystems of size $2$ (crosses). 
The different colors denote entanglement of the following subsystems with the rest of the chain: $\{1, 2\}$ (black), $\{1, 3\}$ (blue), $\{1, 16\}$ (red), and $\{12, 14\}$ (green).
When maximized over all possible subsystems containing two sites (black squares), the entanglement entropy lies close to the theoretical upper bound \eqref{eq:upperBound} (blue dashed line). 
The thin black dashed line is the bound for a subsystem of length $l=1$. 
{\bf Bottom:} The optimized entanglement witness shows a qualitatively similar behavior as the entanglement entropy $S^\mathrm{end}_2$ of typical subsystems. Compared to the maximized $S^\mathrm{end}_2$, however, $W^{\rm end}$ shows a larger number of instances with small entanglement, indicating that it does not capture all entanglement present in the system. 
As a result, correlations between success probability and $W^{\rm end}$ are washed out. 
\label{fig:Fig3}
}
\end{figure}

For computing the dynamics in closed systems, we employ exact diagonalization with Krylov time evolution~\cite{park86} for $N=16$ spins, and with about 100 disorder realizations of $h_i^z$ for each value of $\ta$. Note, that the focus of this work is the entanglement during the optimization with restricts the number of spins to relatively small sizes. 

Let us first illustrate the dynamics of the entanglement in typical sweeps. Figure~\ref{fig:entanglementDynamics} depicts the entanglement dynamics and success probability of a few instances of randomly chosen magnetic fields $h_i^z$. In these examples, the annealing time is fixed to a moderate value $\ta=500/J$, such that for most instances the sweep is nearly adiabatic (here, as in what follows, we set $\hbar$ to unity). Starting from the separable initial state, entanglement gradually increases up to a maximum around $t/\ta \approx 0.6 -0.7$, i.e., in the vicinity of the equilibrium magnetic phase transition of the homogeneous system. 
During the final stages of the protocol, the quantum fluctuations introduced via the transverse fields $h^x \sum_i \sigma_i^{x}$ become weaker; the entanglement decays while simultaneously the success probability increases. 
There exist, however, particular ``hard'' instances where the system is not capable to disentangle, and thus ends up in a superposition of various eigenstates with a lower success probability. 
The optimal local entanglement witness, which is not an entanglement \emph{measure}, shows stronger oscillations than the other observables, which results in a peculiar additional peak in the hard instance of Fig.~\ref{fig:entanglementDynamics}.

\subsubsection{Final-state entanglement}

We will now quantitatively analyze these observations, starting with the final entanglement at $t=\ta$. Figure~\ref{fig:Fig3}, upper row, depicts the final entanglement $S^\mathrm{end}_2\equiv S^\mathrm{max}_2(\ta)$ as a function of $P$ for various annealing times $\ta$. The smaller $\ta$ the more excitations are created during the sweep.  The result is an effective heating that leads to an increased entanglement entropy at the end of the protocol. At the same time, the generated excitations induce a substantial loss of success probability. 

At small annealing times, the qubit pair with maximal entanglement entropy approaches $S^{\rm end}_2 = 2\log 2$, the von Neumann entropy of an infinite-temperature state. 

For larger annealing times $\ta=500/J$, on the other hand, the maximally entangled qubit pair approaches the bound in Eq.~\eqref{eq:upperBound} for many instances, implying that the system achieves the maximally possible success probability given the entropy of the state. Thus, in this regime it is possible to estimate the success probability, an exponentially hard problem, from purely local measurements involving only resources scaling polynomially with system size,  as discussed in Sec.~\ref{sec:upperBound}. But notice that even for these moderate annealing times there are still hard instances with a small success probability and an associated large final entanglement. 

In Fig.~\ref{fig:Fig3}, lower row, we compare the optimal local entanglement witness with the entanglement entropy. The witness $W^{\rm end}$ reproduces the features of the entanglement entropy of typical subsystems but not of the maximized entanglement entropy $S_2^{\rm end}$. 
For fast sweeps, the optimal witness in the final state approaches $W\approx 1/2$, which corresponds to its infinite-temperature value, suggesting a strong heating of the system. 
Increasing the sweep time reduces the final entanglement for most instances. These findings are analogous to the behavior of the entanglement entropy. 

\begin{figure}[t]
\centering
\includegraphics[width=\columnwidth]{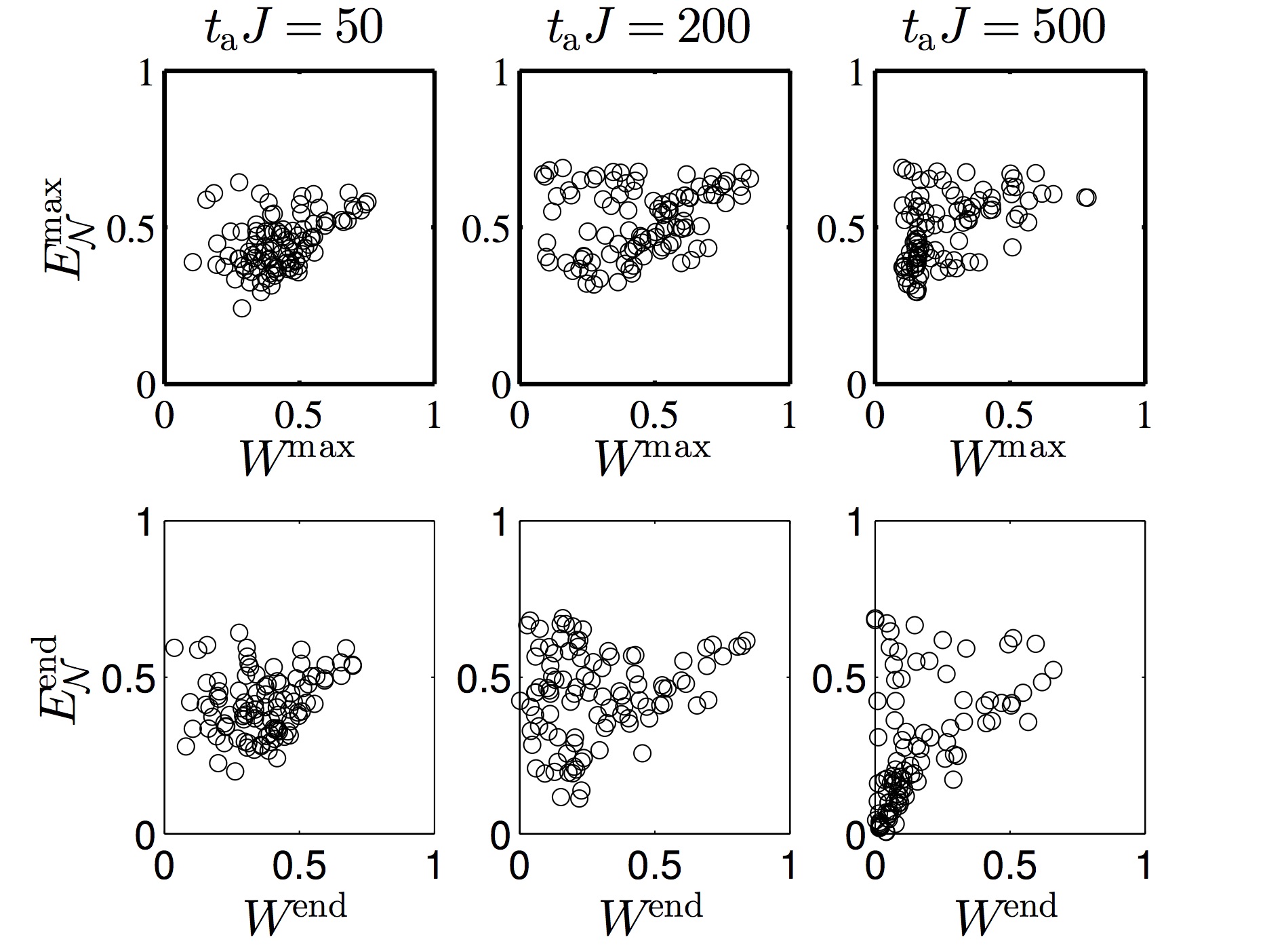}
\caption{
Logarithmic negativity $E_\mN$ compared to the optimized entanglement witness $W$, for different annealing times $\ta$ in a closed system of $N=16$ sites. 
{\bf Top:} Comparison of maximal values during the annealing procedure. {\bf Bottom:} final values. 
$W$ provides a rough lower bound for $E_\mN$. While the tendency for these entanglement observables, is to become smaller with increasing $\ta$, in particular for the final values, correlations to the success probability are difficult to find: for all considered annealing times, both quantities have a broad available range of maximal and final values.  
}
\label{fig:FigNegativityVsWitness}
\end{figure}

For the logarithmic negativity, we find a behavior that is qualitatively similar to the optimal witness. As seen in Fig.~\ref{fig:FigNegativityVsWitness}, the optimal witness provides a rough lower bound for the logarithmic negativity: a large $W$ results in a large $E_\mN$, but not necessarily vice versa. Although the entanglement witness $W^\mathrm{end}$, therefore, can provide a useful operational measure to evaluate the entanglement in the annealing procedure, it shows quantitative differences to more complex entanglement measures. Its shortcomings are clearly seen in some instances that fail to disentangle the state towards the end of the time evolution, where we find strong oscillations in the witness (see \Fref{fig:entanglementDynamics}), which, however, are not accompanied by a further feature in either the entanglement entropy or the logarithmic negativity. 

\begin{figure}[t]
\centering
\includegraphics[width=\columnwidth]{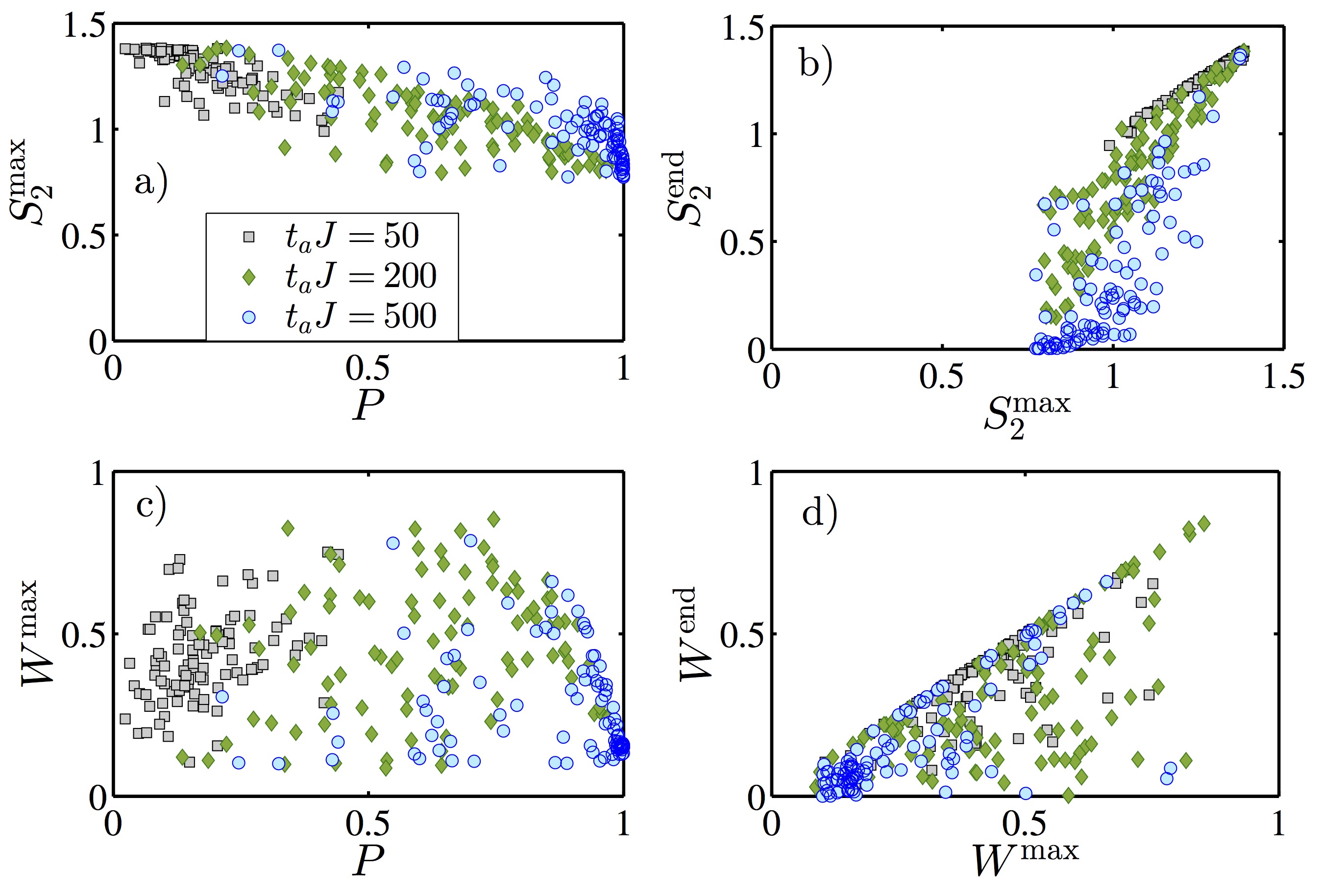}
\caption{
Maximal entanglement reached during the optimization algorithm in closed systems of $N=16$ sites for different annealing times $t_a J = 50$ (black squares), $200$ (green diamonds) and $500$ (blue lozenges).
(a) Entanglement entropy between two sites and the rest of the system, maximized over all pairs of sites, and (c) optimized entanglement witness, both as a function of success probability. 
The measured entanglement occupies a broad range, making it difficult to find correlations to the success probability. 
The maximal entanglement entropy shows only a slight anticorrelation with the success probability $P$, which is inherited from the bound~\eqref{eq:upperBound} and the fact that the final-state entropy for unsuccessful sweeps remains close to the maximal one [panel (b)]. This behavior is less prominent for the entanglement witness [panel (d)]. 
\label{fig:Fig4}
}
\end{figure}

\subsubsection{Maximal entanglement}

From the entanglement entropy, and in particular the bound given in Eq.~\eqref{eq:upperBound}, we found a small entanglement in the final state is crucial for the efficiency. However, the maximal entanglement achieved during the sweep does not seem to be much correlated with the success probability, as can be seen in Fig.~\ref{fig:Fig4}, where we display the behavior of the entanglement entropy $S_2^{\max}=\max_t \left[ S_2^{\rm max}(t) \right]$ (again maximized over all subblocks with $l=2$) as well as the witness $W^{\max}=\max_t W(t)$. 

For large annealing times, the maximal entanglement entropy and witness are spread over a large range of values. This makes it difficult to draw a correlations with the success probability. 
For faster sweeps, there is a slight anticorrelation between $S_2^{\rm max}$ and $P$, but this derives from the bound \eqref{eq:upperBound}, since then the maximal entanglement is retained at the end of the protocol. This can be appreciated in panels (b) and (d) of Fig.~\ref{fig:Fig4}, where it is seen that, for fast annealing, the entanglement at the end is similar to the maximal entanglement. In these figures, one can also see that for slower sweeps, i.e., for more successful instances, the final entanglement is considerably lower than the one attained at the maximum. The logarithmic negativity presents a qualitatively similar behavior (not shown). 

In conclusion, we find no useful correlations between maximal entanglement during the sweep and success probability. In particular, instances that experience a large entanglement during the sweep do not show a larger success probability---contrary to what one might expect when considering the popular picture of spin configurations that quantum tunnel through energy barriers. 

\subsection{Open Quantum Systems}

In an ideal, closed system, the only mechanism to lower the success probability is non-adiabaticity due to finite sweeping times. However, in many current implementations of AQO, the efficiency is also reduced by decoherence in the system. In particular, in the D-wave device, decoherence is orders of magnitude faster than annealing times. Trapped ions, on the other hand, operate in a regime where, as we have shown in Sec.~\ref{sec:trappedIons}, decoherence can be systematically tuned. This ability allows one to study the complex interplay between non-adiabaticity and decoherence in AQO.  In this section, we numerically analyze entanglement witness $W$, negativity $E_\mN$ and the Fisher information $F_Q$ in the AQO protocol. We present numerical data from a relatively small system with $N=6$ sites where considerable statistics is achievable also for numerically exact open-system calculations. In current experiments, the number of ions can go up to $N=20$ \cite{Monroe2013}, which represents an example of a quantum simulator reaching system sizes that cannot be sampled numerically. 

\begin{figure}[h]
\centering
\includegraphics[width=\columnwidth]{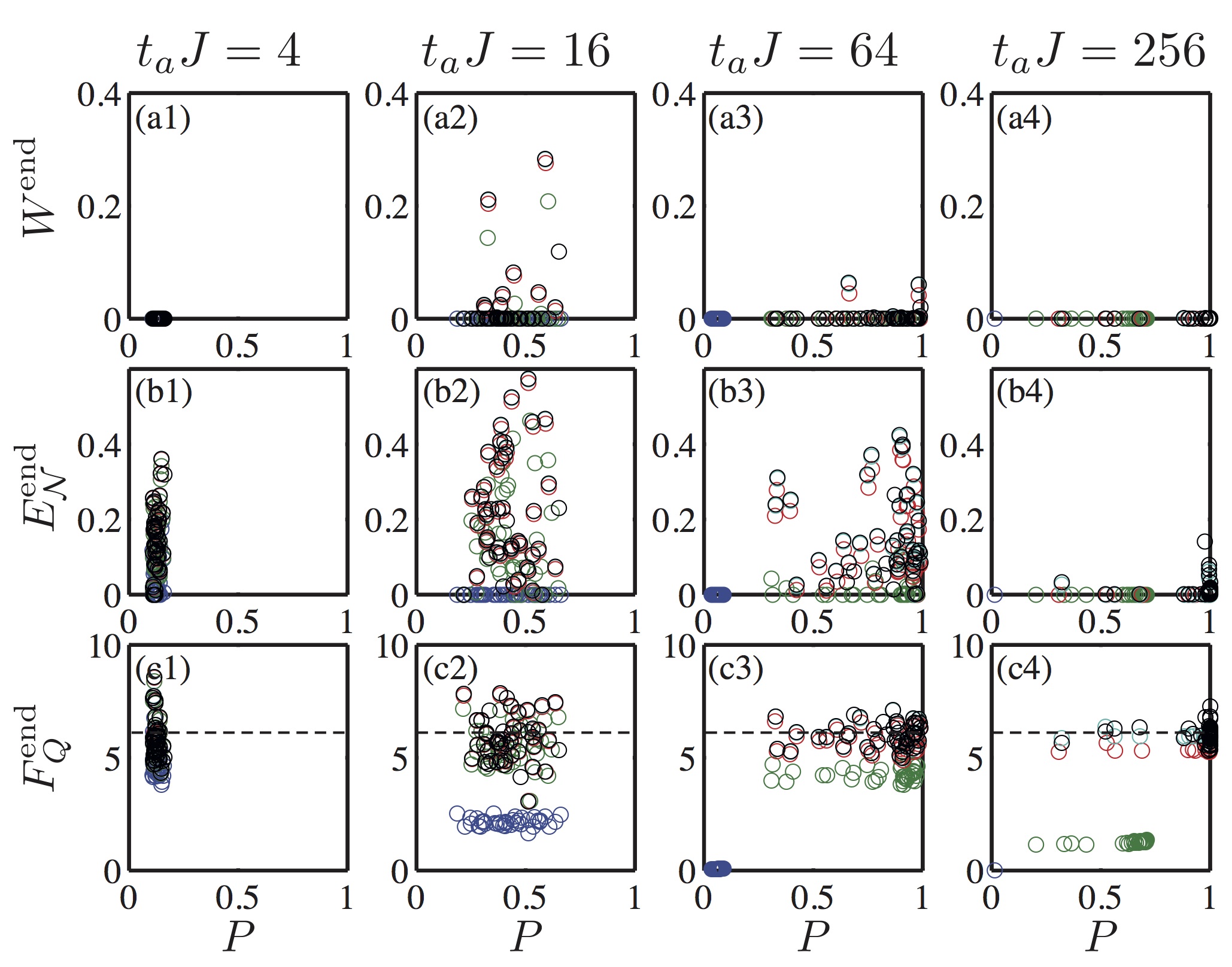}
\caption{
Scatter plot of success probability $P$ against the witness (top), logarithmic negativity (middle), and Fisher information (bottom) for $N=6$ spins. 
Colors indicate the decoherence rate, ranging from $\Gamma=0$ (black), $\Gamma=0.0001J$ (red), $\Gamma=0.001J$ (green), to $\Gamma=0.01J$ (blue). 
For small $\ta$, the final state is always close to an equal superposition of all eigenstates with a small success probability. Fisher information and logarithmic negativity show some residual entanglement. 
At large $\ta$, the success probability decreases with increasing decoherence rate, which also suppresses logarithmic negativity and Fisher information. The dashed line marks $F_Q=N$, above which the system hosts entanglement. 
}
\label{fig:Fig5}
\end{figure}

We compare various annealing times $t_a$, ranging from $\ta=4/J$ to $\ta=256/J$, and decoherence rates ranging from $\Gamma=0$ to $\Gamma=0.01 J$, depicted in Fig.~\ref{fig:Fig5}. For fast sweeps ($\ta=4/J$), the results are independent of $\Gamma$, as can be expected since the considered rates are too low to have any effect in this case. In this regime, the final state remains close to the initial equal superposition of all eigenstates of the final Hamiltonian, and the success probability approaches the limit of $P = 1/2^N$, independent of the rate of decoherence. The entanglement in the final state is then rather uncorrelated from the success probability, which is always low. 

In the opposite extreme of slow sweeps ($\ta=256/J$), the success probability $P$ is governed almost solely by the decoherence rate, and, contrary to the case of closed systems, has little relation to the final-state entanglement. In fact, in this case of extremely slow sweeps, only at small decoherence does any entanglement as measured through $E_\mN$ and $F_Q$ survive, while the witness $W$ vanishes for all studied instances. 
In the regime of intermediate annealing times, for small decoherence rates the success probability and entanglement are related in a similar fashion as in the fully coherent case, but a large decoherence suppresses both entanglement and success probability. 

The optimal witness is rather sensitive to decoherence and does not allow one to obtain detailed information about the process. In contrast, the Fisher information shows rich features and offers additional insight into the nature of the entanglement of the final state. The cases of slow sweeps ($\ta = 256 /J$) and small decoherence rates are particularly interesting for the AQO protocol, since these allow for a large success probability. In these cases, we find that the Fisher information approaches $F_Q \approx N$, which indicates entanglement with $k\approx 1$, i.e., between individual particles. For faster annealing, $F_Q$ can increase slightly, but still remains much below $F_Q=N^2$. We find a similar behavior for the maximum of $F_Q$ assumed during the AQO sweep. This means entanglement between large numbers of particles as measured through $F_Q$ is rare for the studied system sizes. 

State-of-the-art trapped-ion experiments have intrinsic decoherence rates on the order of $\Gamma\sim 0.1 J$, but predominantly only in one quadrature \cite{Jurcevic2014}. We have performed numerical tests also for dephasing only and found an improvement of a factor of about 2 compared to the results presented in Fig.~\ref{fig:Fig5}, where we considered the worst case of equal noise in $\sigma^x$ and $\sigma^z$. Thus, for a systematic study of the influence of noise on AQO in such setups, further improvements in terms of decoherence still seem necessary. Also, as is usual with AQO protocols, larger problem instances will show smaller gaps and thus require larger annealing times to achieve similar degrees of success probability. On the other hand, non-linear ramps may improve the performance of the AQO protocol \cite{Jeremie2002,mandra2014}.

\section{Conclusions\label{sec:conclusions}}

In this work, we presented a feasible way to study the interrelation between entanglement, non-adiabaticity, and decoherence on small scales in a well-controlled laboratory setting. In particular, we proposed several NP-complete problems that can be implemented with straightforward extensions of existing trapped-ion architectures. 
We described how fluctuating synthetic fields allow to engineer well-controlled, artificial decoherence sources that take advantage of the low levels of natural decoherence. Further, we introduced an entanglement witness that is simple to measure in a scalable way, yet reproduces some of the main features of more complex entanglement measures, such as the entanglement entropy and the logarithmic negativity. 

Using extensive numerically exact studies on small closed and open systems, we analyzed the dynamics of entanglement during adiabatic quantum optimization. 
We found that, contrary to popular assumption, a large entanglement during the optimization has little significance for its success probability. 
On the other hand, in clean systems, a large final entanglement after not too fast sweeps indicates that the sweep generated a superposition state as its outcome, with decreased probability to arrive in the separable ground state that solves the optimization problem. 
Decoherence diminishes this anticorrelation between final-state entanglement and success probability. 

Thus, it seems that the simple presence or absence of entanglement does not allow to conclude on the efficiency of a machine performing adiabatic quantum optimization---at least in the case of the entanglement quantifiers and the systems considered in this article. It will be interesting to study whether such correlations appear for larger systems, optimized ramps, or other optimization problems. 
More work is clearly needed to understand if and how entanglement may sign responsible for a quantum speedup of adiabatic quantum optimization.

\emph{Acknowledgements.---}We acknowledge fruitful discussions with Andreas L\"auchli and Peter Zoller. 
This work was supported by the EU integrated project SIQS, ERC synergy grant UQUAM, the Austrian Science Fund (FWF) through the SFB FoQuS (FWF Project No.\ F4018-N23 and F4006-N16), and the Austrian Ministry of Science BMWF as part of the UniInfrastrukturprogramm of the Research Platform Scientific Computing at the University of Innsbruck. WL acknowledges support by the Austrian Science Fund (FWF): P 25454-N27 and MH acknowledges support by the Deutsche Akademie der Naturforscher Leopoldina under grant number LPDS 2013-07.  
Bloch spheres in Fig.~\ref{fig:fig1} and master-equation calculations for open systems used the Qutip libraries \cite{johansson2013qutip}.

\appendix

\section{Generation of noisy transverse Ising models in trapped-ion setups}

The purpose of this Appendix is to identify the assumptions about energy scales entering the derivations of Eqs.~\eqref{eq:HIsingIons} and~\eqref{eq:ME_general} of the main text. We will state them along the way and summarize them in a final section, where we will also provide comparison to the experimental state of art. 

\subsection{Generating Ising interactions using trapped ions} 
\label{sec:AppendixIsingInteractionsInIons}
In trapped-ion experiments, pseudo-spin degrees of freedom can be quantum simulated by choosing two internal electronic states as $\ket{\uparrow}$ and $\ket{\downarrow}$ of a Pauli-spin operator $\tilde\sigma$. 
The energy splitting between these spin states is $\hbar \omega_0$.  
When several ions are confined in an electromagnetic trap, they form---due to their mutual Coulomb repulsion---a regular Wigner-like crystal with collective vibrational modes. 
In a linear trap, the three spatial directions decouple, and the collective ion vibrations are described by three sets of $N$ independent harmonic oscillators, $H_{\rm{ph}}=\sum_{\nu=x,y,z} H_{\rm{ph}}^\nu=\sum_\nu \sum_{q=1}^N \hbar \omega_q^\nu \hat{a}^\dagger_{\nu,q} \hat{a}_{\nu,q}$. Here, $\hat{a}_{\nu,q}^\dagger$ ($\hat{a}_{\nu,q}$) creates (annihilates) a phonon in mode $q$ of spatial direction $\nu$. 
Since these collective modes are extended over all ions, one can use them as a bus to transmit Ising interactions between different spins. 

A convenient method to do this is by coupling the spins and phonons in a M{\o}lmer--S{\o}rensen-type configuration \cite{Sorensen1999a}. 
In this scheme, two laser beams (labeled $l=1$ and $2$) propagate in the same direction, which for the moment we take to be the radial direction $x$. 
Since coupling to the orthogonal vibrational modes can be neglected, we drop the index $\nu=x$  in the following when denoting the phonons. 
The lasers are detuned symmetrically with respect to the vibrational sidebands, i.e., 
$\omega_{l=1}=\omega_0+\omega_{x}+\delta$ and $\omega_{l=2}=\omega_0-\omega_{x}-\delta$, where $\omega_{x}$ is the trap frequency in $x$ direction. 
The coupling to ion $i$ is then described by the Hamiltonian 
\begin{equation}
\label{eq:HL}
H_{\rm L} = \sum_{l=1,2} \sum_i \frac{\hbar|\Omega_i|}{2} (\tilde\sigma_i^+ +\tilde\sigma_i^-)\cos(k x_i-\omega_l t+\phi_l)\,. 
\end{equation} 
Here, $\phi_l$ are the phases of the lasers, which we choose as $\phi_1=0$, $\phi_2=\pi$. We neglect differences in the laser wave number $k$. Further, $|\Omega_i|$ are the absolute values of the laser Rabi frequencies, which we assume equal for the two beams, but which can be different for different ions $i=1\dots N$. The tunability of the latter, to our knowledge so far only considered in Ref.~\cite{Hauke2013b}, is a crucial ingredient for some of the NP-complete models discussed in Table~\ref{tab:NPcompleteProblems}. 

The coupling to the vibrational modes appears through the positions $x_i$, which can be expanded as 
$k x_i=\sum_{q=1}^N \eta_{iq}  {a}_q^\dagger +\mathrm{h.c.}$, where $\eta_{iq}={\mathcal M_{iq}} k/\sqrt{2 M \omega_q/\hbar}$ is the Lamb--Dicke parameter with $M$ the ion mass. The matrices $\mathcal{M}_{i q}$, obtained by diagonalizing the elasticity matrix of the ion crystal, transform localized ion vibrations into phonon normal modes. 
In typical trapped-ion experiments, $\eta_{iq}$ is a small parameter  \cite{Wineland1992}, allowing the expansion of the exponentials in Eq.~(\ref{eq:HL}) to first order. 
Before writing down the resulting expanded Hamiltonian, we change into an interaction picture with respect to $\sum_i \hbar\omega_0 \tilde{\sigma}_i^z+\sum_q (\omega_x+\delta)\hat{a}_q^\dagger \hat{a}_q$, and exploit the hierarchy $\omega_0\gg\omega_x+\delta \gg |\Omega_i|$ to make a standard rotating-wave approximation, yielding 
\begin{equation}
\label{eq:HLI}
H_{\rm{I}} = \sum_q \hbar\delta_q \hat{a}_q^\dagger \hat{a}_q + \sum_i  \frac{\hbar|\Omega_i|}{2}  \sum_q i \eta_{iq} \tilde\sigma_i^x(\hat{a}_q^\dagger-\hat{a}_q)\,,
\end{equation} 
where $\delta_q\equiv \omega_q-\omega_x-\delta$ describes the detuning relative to the sidebands. 
To make better contact to the discussion in the main text, before proceeding we rotate the spins by $\pi/2$ about the $y$ axis of their Bloch sphere, which sends $\tilde\sigma_i^x\to \sigma_i^z$ and $\tilde\sigma_i^z\to -\sigma_i^x$. 

The spin--phonon coupling contained in Hamiltonian (\ref{eq:HLI}) transmits an interaction between spins, as can be revealed, e.g., by considering the canonical transformation 
$H_{\mathrm{I}}^\prime=\ue^{\mathcal S} H_{\rm{I}} \ue^{-\mathcal S}$, with 
${\mathcal S}=\sum_{iq} g_{iq}^z (\hat{a}^\dagger_q+\hat{a}_q)\sigma_i^z$, where 
$g_{iq}^z=\frac{i}{2}\frac{|\Omega_i|\eta_{iq}}{\delta_q}$. 
For $|\Omega_i|\eta_{iq}/{\delta_q}\ll 1$, we can expand $H_{\mathrm{I}}^\prime$ up to second order, and---neglecting constant terms---we obtain the promised Ising interactions
\begin{equation}
\label{eq:HpureIsing}
H_{\mathrm{I}}^\prime= \sum_q \delta_q \hat{a}_q^\dagger \hat{a}_q + \sum_{i\neq j} J_{ij} \sigma_i^z \sigma_j^z\,, 
\end{equation}
with 
\begin{equation}
\label{eq:Jij}
J_{ij} = - \hbar\frac{|\Omega_i| |\Omega_j|}{4}\sum_q \frac{\eta_{iq} \eta_{jq}}{\delta_q}\,.
\end{equation}
In this form of $H_{\mathrm{I}}^\prime$, spins and phonons are effectively decoupled, allowing consideration of the spin-only model $H_{\mathrm{Ising}}$ as given in Eq.~(\ref{eq:HIsingIons}) of the main text. 
Via the detuning $\delta_q$, the exponent $\alpha$ can be tuned in the range from 0 to 3 \cite{Britton2012,Jurcevic2014}. 
Equations (\ref{eq:HpureIsing}) and (\ref{eq:Jij}) are (up to small corrections) a standard form of Ising couplings describing current trapped-ion experiments \cite{Britton2012,Richerme2014,Jurcevic2014}, but with the additional freedom of choosing local Rabi frequencies made explicit. 

\subsection{Effective longitudinal and transverse fields} 
To implement the NP-complete models given in the main text, we require also longitudinal-field terms $H_{\parallel}=\sum_i h_i^z \sigma_i^z$. 
In the M{\o}lmer--S{\o}rensen scheme, these terms can effectively be implemented by resonantly driving the spin transition at $\omega_0$ in phase with the laser beams. 
Since $\left[\sigma_i^z,\mathcal{S}\right]=0$, we have that $H_{\parallel}$ is unaffected by the canonical transformation and it may simply be added to the final Hamiltonian. 

The case is a bit different when adding the transverse-field terms $H_{\perp}=\sum_i h_i^x \sigma_i^x$. 
In experiment, these terms can be obtained, e.g., by shifting the detuning of both M{\o}lmer--S{\o}rensen beams with respect to the pseudospin transition, or by an additional beam tuned in resonance to the pseudospin transition but with a phase shift of $\pi/2$. 
Such transverse fields generate an additional coupling of spins to phonons. 
The question of the correct parameter regimes where this coupling can be neglected is a difficult subject, and has been studied in detail in Ref.~\cite{Wang2012}. 
We give here an estimate for when additional phonon heating due to effective fields can be neglected. 
For small effective fields, one can incorporate the fields by modifying the canonical transformation to 
$\mathcal{S}=\sum_{\nu=x,y,z} \sum_{iq} g_{iq}^\nu(\gamma_\nu \hat{a}_{iq}^\dagger + \gamma_\nu^\star \hat{a}_{iq}) \sigma_i^\nu$, 
where 
$\gamma_z=\gamma_x=1$, $\gamma_y=i$. 
To leading order, 
$g_{iq}^z=\frac{i}{2}\frac{|\Omega_i|\eta_{iq}}{\delta_q}$ remains unchanged.
However, we also have the additional, lower-order terms 
$g_{iq}^y=g_{iq}^z\frac{h_i^z}{\hbar\delta_q/2}$
and  
$g_{iq}^x=g_{iq}^z\frac{h_i^x h_i^z}{(\hbar\delta_q/2)^2}$. 
Using this modified form of the canonical transformation, one obtains, up to second order in $(h_i^x,h_i^z,\hbar\eta_{iq}|\Omega_i|)/\hbar\delta_q$,  the unchanged effective Hamiltonian
$H_{\mathrm{Ising}} + H_{\parallel}+H_{\perp}$, i.e., as long as $h_i^{x,z}\ll\hbar\delta_q$, we can neglect additional corrections due to the synthetic magnetic fields.

\subsection{Noise engineering in trapped ions\label{sec:AppendixNoiseInIons}} 
The ion system by itself is subject to a certain decoherence due, e.g., to (real) magnetic-field fluctuations or spontaneous decay of the upper pseudospin level. 
However, it is possible to also engineer a desired noise source, and thus study the effect of designed decoherence on the adiabatic optimization, as we describe now. 
A scheme to engineer noise in a controlled fashion has also been discussed and implemented in Ref.~\cite{Soare2014}.

The basic idea is to subject the spins to fluctuating terms, generated in the same manner as for the static effective fields, but with time-dependent fluctuating amplitudes. 
The full Hamiltonian in presence of such classical dephasing is modified to 
\begin{equation}
	\label{eq:Hnoise}
	H  = H_{\mathrm{I}} + H_{\parallel}+H_{\perp} + \sum_i \left[\xi_i^x(t) \sigma_i^x + \xi_i^z(t) \sigma_i^z\right]\,.
\end{equation}
We take $\xi_i^\nu(t)$, $\nu=x,z$, to be independent Ornstein--Uhlenbeck processes, $\llangle\, \xi_i^\nu(t) \xi_j^\mu(0) \,\rrangle=\hbar\Gamma \delta_{\nu,\mu} \delta_{i,j} \hbar b \ue^{-b |t|}$, where $\llangle\, \cdot \,\rrangle$ denotes average over noise realizations. 
For simplicity, we assume equal rates $\Gamma$ and bandwidths $b$ for both the $\xi_i^x(t)$ and $\xi_i^z(t)$ noise. 

At times $t\ll (\eta_{iq} |\Omega_i|)^{-1}$, we can consider spins and phonons as uncoupled. The equation of motion for the phonon density matrix $\rho_{\rm ph}$ is then $\dot\rho_{\rm ph}=-\frac{i}{\hbar} [H_{\rm ph},\rho_{\rm ph}]$ and for the spins (up to second order in $\xi_i^\nu$)
\begin{eqnarray}
\dot\rho(t)&=&-\frac{i}{\hbar} [H_{\parallel}+H_{\perp},\rho] \\
	& &-\frac{1}{\hbar^2} \int_0^t \ud t^\prime \sum_{i,j}\sum_{\nu,\mu} \xi_i^\nu(t) \xi_j^{\mu}(t^\prime) \left[\sigma_i^\nu,\left[\sigma_j^{\mu},\rho(t^\prime)\right]\right]\,.  \nonumber
\end{eqnarray}
In the limit where $b$ is much larger than all other relevant time scales, $b\gg h_i^\nu/\hbar\,, \eta_{iq} |\Omega_i|$, putting the upper integration limit to infinity and averaging over noise realizations leaves us with the master equation for the spins
\begin{equation}
\llangle\dot\rho\rrangle(t)=-\frac{i}{\hbar} [H_{\parallel}+H_{\perp},\llangle\rho\rrangle] 
-\Gamma \sum_{i,\nu}   \left[\sigma_i^\nu,\left[\sigma_i^{\nu},\llangle \rho(t)\rrangle\right]\right]\,.
\end{equation}
From here, one can add the spin--phonon coupling and use the same canonical transformation as in the Hamiltonian case to arrive at the master Eq.~\eqref{eq:ME_general}, where we omitted the double angle brackets for simplicity. 

At larger times, $t= \mathcal{O}\left((\eta_{iq} |\Omega_i|)^{-1}\right)$, spins and phonons get coupled, and we have to consider the master equation for the density matrix $\rho_{\mathrm{f}}$ of the full system, 
\begin{equation}
\label{eq:masterEq}
\llangle\dot\rho_{\mathrm{f}}\rrangle(t)=-\frac{i}{\hbar} [H_{\rm I}+H_{\parallel}+H_{\perp},\llangle\rho_{\mathrm{f}}\rrangle] 
-\Gamma \sum_{i,\nu}   \left[\sigma_i^\nu,\left[\sigma_i^{\nu},\llangle \rho_{\mathrm{f}}\rrangle\right]\right]\,.
\end{equation}
Using the above canonical transformation on the non-hermitian term introduces spin--phonon couplings at the order of $\Gamma \eta_{iq}|\Omega_i|/\delta_q$. These spin--phonon couplings introduce additional decoherence for the spin subsystem, but since $\eta_{iq}|\Omega_i|\ll \delta_q$ they are much smaller than the primary decoherence term included in Eq.~(\ref{eq:masterEq}). Therefore, we can neglect them, and arrive at the master equation quoted in the main text, Eq.~(\ref{eq:ME_general}). 

\subsection{Hierarchy of energy scales and rates}
In the derivation of Eq.~(\ref{eq:ME_general}), several subleading energy scales have been neglected. We now discuss the validity of the underlying assumptions in view of experimental parameters such as used in the experiments of Refs.~\cite{Richerme2014,Jurcevic2014}. 
For the validity of the rotating wave approximation leading to Eq.~(\ref{eq:HLI}), one requires $\omega_0\gg\omega_x+\delta \gg |\Omega_i|$. Typically, $\omega_x$ is $\mathcal{O}(1\,\rm{MHz})$, and $\delta$ and $|\Omega_i|$ are chosen on the orders $\mathcal{O}(10-100\,\rm{kHz})$ and $\mathcal{O}(100\,\rm{kHz})$, respectively. 
The pseudospin energy splitting is on the order of $\mathcal{O}(10\,\rm{GHz})$ for hyperfine transitions and can attain optical frequencies. Therefore, this first set of inequalities is safely fulfilled. 

Further, the expansion of the canonical transformation $\ue^{\mathcal{S}}$ requires $h_i^{x,z}/\hbar, |\Omega_i| \eta_{iq}\ll\delta_q$. 
Typical values are $\eta_{iq}\approx 0.06$, $\delta_q=\omega_q-\omega_x-\delta=\mathcal{O}(10-100\,\rm{kHz})$, and 
$|\Omega_i| = \mathcal{O}(100\,\rm{kHz})$. 
Considering the example of the Coulomb glass, in Ref.~\cite{Jurcevic2014} it has been shown that for $\delta_q\approx 40\,\rm{kHz}$, the interactions in a chain of 15 ions can be well approximated by a power law with exponent $\alpha\approx 1$. 
In this case, a good choice of the Rabi frequency could be $|\Omega_i| \approx 107\,\rm{kHz}$, yielding the expansion parameter $|\Omega_i|\eta_{iq}/\delta_q\approx 0.16$.
The resulting interaction strength would be $J/\hbar \approx 1\,\rm{kHz}$. The transverse field at the beginning of the annealing protocol, of order $J$, does therefore fulfill $h^x/\hbar\approx 1\,{\rm kHz} \ll \delta_{q}$. 
Let us stress that these are values that are already used in the laboratory, which may be open to improvement in the future. 

Spin--phonon couplings introduced when engineering decoherence can be neglected when $b,\delta_q\gg \eta_{iq}|\Omega_i|$. 
The second inequality was already required for the expansion of the canonical transformation. 
For the above parameter values, the first inequality implies $b\gg 6\, {\rm kHz}$, which also ensures validity of the white-noise assumption $b\gg h_i^\nu$. 
Such a fast and spatially uncorrelated modulation could be generated, e.g., via acousto-optic deflectors (AODs) as are employed to achieve single-ion addressing \cite{Jurcevic2014}. Since the response time of AODs can be less than $10\,\mu \rm{s}$, it is possible to let the intensity incident on different ions fluctuate independently and much faster than $\eta_{iq}|\Omega_i|,h_i^\nu$. 

A fast fluctuation (large $b$) ensures that the noise-generating fields do not introduce additional spin--phonon couplings in higher orders of the canonical transformation as the static synthetic fields do. For slower fluctuations, however, phonons could be excited if the typical fluctuation amplitude does not fulfil the condition analog to the static counterpart, $\sqrt{\llangle\, \xi_i^\nu(0) \xi_i^\nu(0) \,\rrangle}/\hbar=\sqrt{\Gamma b}\ll \delta_q$. For the above parameters and the small noise rates considered in this work,  $\Gamma \leq 0.01 J/\hbar$, however, this condition is safely satisfied. 

We can summarize the required hierarchy of energy scales and rates as $\delta_q,b \gg \eta_{iq}|\Omega_i|,h_i^{\nu}/\hbar$ and $\delta_q\gg \sqrt{\Gamma b}$.
Let us finally stress that incomplete fulfillment of these hierarchies may actually not constitute an issue, since we are interested in analyzing the interplay between non-adiabaticity and decoherence, i.e., we explicitly desire these detrimental effect, only with the restriction that they should be tunable. For example, increasing the Rabi frequency to obtain a larger $J$ improves adiabaticity, but may simultaneously introduce additional decoherence due to excitation of phonons.

\bibliographystyle{apsrev}

\end{document}